# Joint Maximum Likelihood Estimation of Carrier and Sampling Frequency Offsets for OFDM systems

Yong-Hwa Kim

Korea Electrotechnology Research Institute (KERI), and University of Science and Technology (UST), Korea. E-mail: yongkim@keri.re.kr

**Abstract**

In orthogonal-frequency division multiplexing (OFDM) systems, carrier and sampling frequency offsets (CFO and SFO, respectively) can destroy the orthogonality of the subcarriers and degrade system performance. In the literature, Nguyen-Le, Le-Ngoc, and Ko proposed a simple maximum-likelihood (ML) scheme using two long training symbols for estimating the initial CFO and SFO of a recursive least-squares (RLS) estimation scheme. However, the results of Nguyen-Le's ML estimation show poor performance relative to the Cramer-Rao bound (CRB). In this paper, we extend Moose's CFO estimation algorithm to joint ML estimation of CFO and SFO using two long training symbols. In particular, we derive CRBs for the mean square errors (MSEs) of CFO and SFO estimation. Simulation results show that the proposed ML scheme provides better performance than Nguyen-Le's ML scheme.

# 1. Introduction

Orthogonal frequency-division multiplexing (OFDM) has received considerable attention for its robustness against frequency-selective fading channels. It is known that the drawback of the OFDM system is the sensitivity of the receiver to oscillator instabilities such as carrier and sampling frequency offsets (CFO and SFO, respectively).

Both CFO and SFO are introduced by a mismatch of the oscillator frequencies between the transmitter and receiver, which causes intercarrier interference (ICI) [1], [2]. The maximum-likelihood (ML) estimation for CFO is derived using the repeated training symbols by Moose [3]. However, the detrimental SFO effect has been omitted in Moose's ML estimation. In [4], Nguyen-Le, Le-Ngoc, and Ko proposed a simple ML estimator for the coarse estimation of initial CFO and SFO values to enhance the performance and convergence of recursive least-squares (RLS) estimation and tracking of the CFO, SFO, and channel impulse response (CIR). However, in terms of the estimation errors, Nguyen-Le's ML estimator shows poor performance compared to the Cramer-Rao bounds (CRBs). Coarse CFO and SFO estimates are used as the initial values for RLS-based joint CFO, SFO, and CIR estimation and tracking. This results in performance degradation of the RLS-based iterative scheme due to initial estimation errors.

In this paper, we describe an extension of Moose's CFO estimation scheme to joint estimation of CFO and SFO for increasing estimation accuracy when repeated training symbols are available. Moreover, we derive a closed form expression for the CRBs associated with the estimation of both CFO and SFO. It is found by simulations that the proposed ML estimator yields an estimation performance superior to that of Nguyen-Le's ML estimator.

## 2. System Model

Let $N$ be the number of subcarriers, $K$ the number of modulated subcarriers, and $N_g$ the number of cyclic prefix (CP). Note that $N-K$ subcarriers at the edges of the spectrum are not used, and the modulated subcarriers can be indexed by numbers ranging from $-K/2$ to $K/2-1$. At the receiver, the mismatch between the local oscillators in the transceiver and Doppler shifts cause a frequency offset $\Delta f$ Hz. The mismatch between sampling clocks in the transceiver causes SFO $\eta$. The received signals are therefore sampled at a rate of $1/T'$, where $T'=(1+\eta)T$ and $T$ is the sampling period at the transmitter. When the preamble signal is composed of $M=2$ long training symbols, the time-domain received signal on the $n$th sample of the $m$th OFDM training symbol becomes [4]

$$r_{m,n} = \frac{e^{j\frac{2\pi}{N}(N_m+n)(1+\eta)\varepsilon}}{\sqrt{N}} \sum_{k=-K/2}^{K/2-1} X_m(k)H(k)e^{j\frac{2\pi k}{N}n(1+\eta)} e^{j\frac{2\pi k}{N}\eta N_m} + w_m(n), \qquad (1)$$

where $m=0,1$; $n=0,1,\cdots,N-1$; $N_m = N_g + m(N+N_g)$; $\varepsilon = \Delta f N T$ is the normalized CFO; $X_m(k)$ denotes the modulated symbol on the $k$th subcarrier; $H(k) = \sum_{l=0}^{L-1} h_l e^{-j(2\pi/N)kl}$ is the frequency-domain channel response; $h_l$ is the $l$th tap of the CIR; $L$ is the number of the channel taps; and $w_m(n)$ is the additive white Gaussian noise (AWGN) with zero mean and variance $\sigma_w^2$.

After the discrete Fourier transform (DFT), the resulting frequency-domain signal $R_m(k)$ for $k \in \mathbb{M}$ from (1) is given as

$$R_m(k) = \delta_{kk} e^{j\frac{2\pi}{N}N_m(k\eta+\varepsilon(1+\eta))} X_m(k)H(k) + ICI_m(k) + W_m(k), \qquad (2)$$

where $\mathbb{M}$ denotes the integer subcarrier index set consisting of $\{-K/2,\cdots,-1,0,1,\cdots,K/2-1\}$; the ICI noise $ICI_m(k)$ is defined as

$$ICI_m(k) = \sum_{i \in \mathbb{M}, i \neq k} \delta_{ki} e^{j\frac{2\pi}{N}N_m(i\eta+\varepsilon(1+\eta))} X_m(i) H(i), \tag{3}$$

and $W_m(k)$ is the frequency response of $w_m(n)$. In (2) and (3), $\delta_{ki}$ is a function of CFO $\varepsilon$ and SFO $\eta$ and is given by

$$\delta_{ki} = \frac{1}{N} \sum_{n=0}^{N-1} e^{j\frac{2\pi}{N}n(i\eta+\varepsilon(1+\eta)+i-k)}. \tag{4}$$

We can then rewrite (2) in a compact matrix format as follows:

$$\mathbf{R}_m = \mathbf{\Omega}_m \mathbf{X}_m \mathbf{H} + \mathbf{ICI}_m + \mathbf{W}_m, \tag{5}$$

$$\mathbf{\Omega}_m = diag\{\Omega_m(-K/2), \Omega_m(-K/2+1), \cdots, \Omega_m(K/2-1)\}, \tag{5a}$$

$$\Omega_m(k) = \delta_{kk} e^{j\frac{2\pi}{N}N_m(k\eta+\varepsilon(1+\eta))}, \tag{5b}$$

$$\mathbf{X}_m = diag\{X_m(-K/2), X_m(-K/2+1), \cdots, X_m(K/2-1)\}, \tag{5c}$$

$$\mathbf{H} = [H(-K/2)\ H(-K/2+1)\ \cdots\ H(K/2-1)]^T, \tag{5d}$$

$$\mathbf{ICI}_m = [ICI_m(-K/2)\ ICI_m(-K/2+1)\ \cdots\ ICI_m(K/2-1)]^T, \tag{5e}$$

$$\mathbf{W}_m = [W_m(-K/2)\ W_m(-K/2+1)\ \cdots\ W_m(K/2-1)]^T. \tag{5f}$$

Here, we define the ICI-plus-noise vector as $\mathbf{V}_m = \mathbf{ICI}_m + \mathbf{W}_m$. The vector $\mathbf{V}_m$ can be approximated by a Gaussian-distributed vector with zero mean and covariance matrix $\sigma_V^2 \mathbf{I}_K$ [5].

## 3. Joint ML synchronization

In this section, a joint ML estimation algorithm of the CFO and SFO is derived as an extension of Moose's CFO estimation algorithm, which was first proposed solely for CFO estimation [3]. For comparison purposes, we reviewed Nguyen-Le's ML estimator for both CFO and SFO [4].

In the following, we present a new joint ML estimation scheme to obtain both the CFO $\varepsilon$ and SFO $\eta$ that maximize the conditional probability density function (pdf) $p(\mathbf{R}_0, \mathbf{R}_1 | \varepsilon, \eta)$. The ML estimates of $\varepsilon$ and $\eta$ can be obtained by

$$\hat{\varepsilon}, \hat{\eta} = \arg\max_{\varepsilon, \eta} p(\mathbf{R}_0, \mathbf{R}_1 | \varepsilon, \eta) = \arg\max_{\varepsilon, \eta} p(\mathbf{R}_1 | \varepsilon, \eta, \mathbf{R}_0) p(\mathbf{R}_0 | \varepsilon, \eta). \tag{6}$$

Assuming that the training symbols are identical (that is, $\mathbf{X}_0 = \mathbf{X}_1$) for easier synchronization [3], we have

$$\mathbf{R}_0 = \mathbf{\Omega}_0 \mathbf{X}_0 \mathbf{H} + \mathbf{V}_0, \tag{7}$$

and

$$\mathbf{R}_1 = \mathbf{\Xi}(\varepsilon, \eta) \mathbf{R}_0 + \mathbf{N}, \tag{8}$$

where

$$\mathbf{\Xi}(\varepsilon, \eta) = diag\left\{ e^{j\frac{2\pi(N+N_g)}{N}\left(-\frac{K}{2}\eta + \varepsilon(1+\eta)\right)}, e^{j\frac{2\pi(N+N_g)}{N}\left(\left(-\frac{K}{2}+1\right)\eta + \varepsilon(1+\eta)\right)}, \cdots, e^{j\frac{2\pi(N+N_g)}{N}\left(\left(\frac{K}{2}-1\right)\eta + \varepsilon(1+\eta)\right)} \right\},$$

and $\mathbf{N} = -\mathbf{\Xi}(\varepsilon, \eta)\mathbf{V}_0 + \mathbf{V}_1$. Assuming that $p(\mathbf{R}_0 | \varepsilon, \eta) = p(\mathbf{R}_0)$ ($\varepsilon$ and $\eta$ give no information about $\mathbf{R}_0$), Eq. (6) reduces to

$$\hat{\varepsilon}, \hat{\eta} = \arg\max_{\varepsilon, \eta} p(\mathbf{R}_1 | \varepsilon, \eta, \mathbf{R}_0), \tag{9}$$

where $p(\mathbf{R}_1 | \varepsilon, \eta, \mathbf{R}_0)$ has a mean of $\mathbf{\Xi}(\varepsilon, \eta)\mathbf{R}_0$ and a covariance of $2\sigma_V^2 \mathbf{I}_K$. The likelihood function $p(\mathbf{R}_1 | \varepsilon, \eta, \mathbf{R}_0)$ can be given by

$$p(\mathbf{R}_1 | \varepsilon, \eta, \mathbf{R}_0) = \frac{1}{(\pi 2\sigma_V^2)} \exp\left\{-\frac{1}{2\sigma_V^2} \|\mathbf{R}_1 - \mathbf{\Xi}(\varepsilon, \eta)\mathbf{R}_0\|^2 \right\}, \tag{10}$$

where $\|\mathbf{x}\|^2 = \mathbf{x}^H \mathbf{x}$. The proposed ML estimates $\hat{\varepsilon}$ and $\hat{\eta}$ for CFO and SFO, respectively, can be obtained by

$$\hat{\varepsilon}, \hat{\eta} = \arg\min_{\varepsilon, \eta} \sum_{k \in \mathbb{M}} \left| R_1(k) - e^{j\frac{2\pi(N+N_g)}{N}(k\eta + \varepsilon(1+\eta))} R_0(k) \right|^2. \tag{11}$$

Exploiting two long training symbols, Nguyen-Le et al. proposed a simple ML estimator for the joint estimation of CFO and SFO [4], which involves minimization of the following cost function

$$\hat{\varepsilon}, \hat{\eta} = \arg\min_{\varepsilon, \eta} \sum_{k \in \mathbb{M}} \left| Y(k) - e^{j\frac{2\pi}{N}(N+N_g)(k\eta+\varepsilon(1+\eta))} \right|^2, \quad (12)$$

where

$$Y(k) = \frac{X_0(k) R_1(k)}{X_1(k) R_0(k)} = e^{j\frac{2\pi}{N}(N+N_g)(k\eta+\varepsilon(1+\eta))} + E(k). \quad (13)$$

As shown in [4], $E(k)$ is defined by

$$E(k) = \frac{X_0(k)(ICI_1(k) + W_1(k)) - X_1(k)(ICI_0(k) + W_0(k)) e^{j\frac{2\pi}{N}(N+N_g)(k\eta+\varepsilon(1+\eta))}}{X_0(k) X_1(k) H(k) \delta_{kk} e^{j\frac{2\pi}{N} N_g (k\eta+\varepsilon(1+\eta))} + X_1(k)(ICI_0(k) + W_0(k))}, \quad (14)$$

and can also be approximated as uncorrelated and Gaussian-distributed. We can then rewrite (13) in a compact matrix format as follows:

$$\mathbf{Y} = \Xi(\varepsilon, \eta) \mathbf{1}_K + \mathbf{E}, \quad (15)$$

where $\mathbf{Y} = [Y(-K/2) \cdots Y(K/2-1)]^T$, $\mathbf{1}_K = [1, 1, \cdots, 1]^T$, and $\mathbf{E} = [E(-K/2) \cdots E(K/2-1)]^T$. Under the assumption that the error vector $\mathbf{E}$ is a Gaussian-distributed vector with zero mean and covariance matrix $\sigma_E^2 \mathbf{I}_K$ [4], the likelihood function $p(\mathbf{Y}|\varepsilon, \eta)$ can be given by

$$p(\mathbf{Y}|\varepsilon, \eta) = \frac{1}{(\pi \sigma_E^2)} \exp\left\{ -\frac{1}{\sigma_E^2} \|\mathbf{Y} - \Xi(\varepsilon, \eta) \mathbf{1}_K\|^2 \right\}. \quad (16)$$

The above estimates of CFO and SFO in (12) are used as the initial estimates for the RLS-based iterative estimation of CFO, SFO, and CIR [4].

Considering the likelihood functions (10) and (16), the variances of the error terms $2\sigma_V^2$ and $\sigma_E^2$ influence the accuracy of the ML estimators, respectively. Fig. 1 presents the noise variances of the error terms $2\sigma_V^2$ and $\sigma_E^2$ with $\varepsilon = 0.212$ and $\eta = 0.000112$ in a Rayleigh fading channel, where $2\sigma_V^2$ and $\sigma_E^2$ can be calculated as $E\left[\|\mathbf{N}\|^2\right]$ and $E\left[\|\mathbf{E}\|^2\right]$, respectively, and the number of multipaths for the Rayleigh fading channel is $L=5$. At all ranges of signal-to-noise ratios (SNRs), the noise variance $2\sigma_V^2$ for (10) shows better performance than the noise variance $\sigma_E^2$ for (16). For example, at SNR=15 dB, the noise variance $2\sigma_V^2$ for (10) is about 13.5 dB smaller than the noise variance $\sigma_E^2$ for (16).

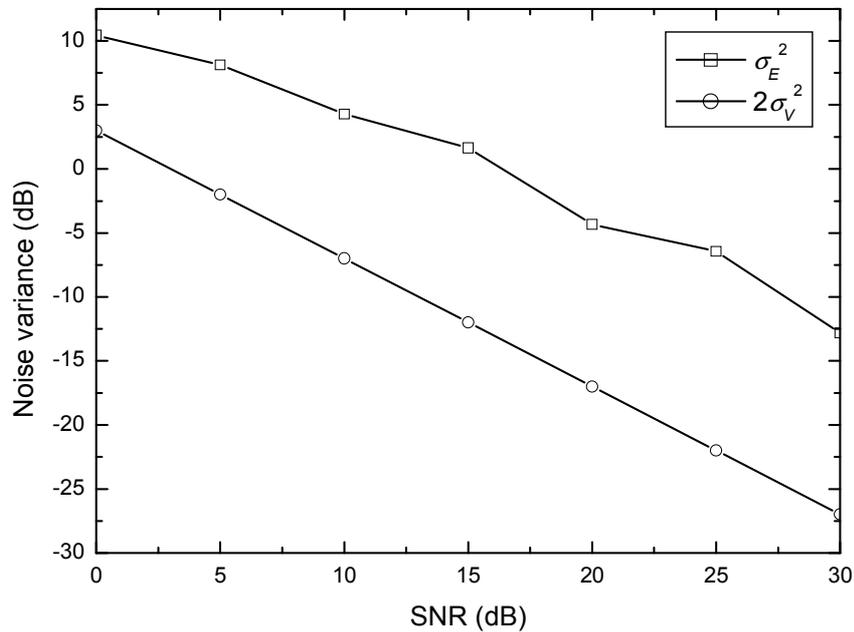

Fig. 1. Noise variance versus SNR in a Rayleigh fading channel where $N=64$, $K=52$, $N_g=16$, $\varepsilon=0.212$, and $\eta=0.000112$

## 4. Cramer-Rao Bound for CFO and SFO estimation

For CRBs in CFO, SFO, and CIR estimation, Nguyen-Le et al. presented a calculation of the Fisher information matrix [4]. Unfortunately, the results presented in that paper had calculation errors [4, Eqs. (28) and (29)]. In this section, CRBs are derived for both CFO and SFO estimators. Discarding terms not dependent on $\varepsilon$ and $\eta$, we can obtain the log-likelihood function of the received signal by

$$\Lambda = -\frac{1}{\sigma_W^2} \sum_{m=0}^{M-1} \sum_{n=0}^{N-1} \left| r_{m,n} - \frac{e^{j\frac{2\pi}{N}(N_m+n)(1+\eta)\varepsilon}}{\sqrt{N}} \sum_{k \in \mathbb{M}} X_m(k) H(k) e^{j\frac{2\pi k}{N}n(1+\eta)} e^{j\frac{2\pi k}{N}\eta N_m} \right|^2. \quad (17)$$

Defining $\varpi_0 = \varepsilon$ and $\varpi_1 = \eta$, we can represent the Fisher information matrix $\mathbf{F}$ as

$$[\mathbf{F}]_{i,j} = -E\left\{\frac{\partial^2 \Lambda}{\partial \varpi_i \partial \varpi_j}\right\} \quad (18)$$

where

$$F_{0,0} = -E\left\{\frac{\partial^2 \Lambda}{\partial \varepsilon \partial \varepsilon}\right\}$$
$$= \frac{2}{\sigma_W^2 N} \sum_{m=0}^{M-1} \sum_{n=0}^{N-1} \left[\frac{2\pi}{N}(1+\eta)(N_m+n)\right]^2 \left| \sum_{k=-K/2}^{K/2-1} X_m(k) H(k) e^{j\frac{2\pi k}{N}n(1+\eta)} e^{j\frac{2\pi k}{N}\eta N_m} \right|^2, \quad (19)$$

$$F_{1,0} = F_{0,1} = -E\left\{\frac{\partial^2 \Lambda}{\partial \eta \partial \varepsilon}\right\}$$
$$= \frac{-2}{\sigma_W^2 N} \sum_{m=0}^{M-1} \sum_{n=0}^{N-1} \mathrm{Re}\left\{j\frac{2\pi}{N}(N_m+n)(\Phi_{m,n} + \Psi_{m,n})\right\} \quad (20)$$

with

$$\Phi_{m,n} = \left(1 + j\frac{2\pi(N_m+n)(1+\eta)\varepsilon}{N}\right) \left| \sum_{k=-K/2}^{K/2-1} X_m(k) H(k) e^{j\frac{2\pi k}{N}n(1+\eta)} e^{j\frac{2\pi k}{N}\eta N_m} \right|^2, \quad (20\text{a})$$

$$\Psi_{m,n} = j\frac{2\pi(N_m+n)}{N}(1+\eta)$$
$$\times \sum_{k=-K/2}^{K/2-1} \sum_{k'=-K/2}^{K/2-1} k' X_m(k') X_m^*(k) H(k') H^*(k) e^{j\frac{2\pi(k'-k)}{N}n} e^{j\frac{2\pi(k'-k)}{N}\eta(N_m+n)}, \quad (20\text{b})$$

and

$$F_{1,1} = -E\left\{\frac{\partial^2 \Lambda}{\partial \eta \partial \eta}\right\} = -\frac{2}{\sigma_W^2 N} \sum_{m=0}^{M-1} \sum_{n=0}^{N-1} \mathrm{Re}\{\Gamma_{m,n} + \Theta_{m,n} + \Pi_{m,n}\}, \quad (21)$$

with

$$\Gamma_{m,n} = -\left(\frac{2\pi}{N}(N_m+n)\right)^2 \varepsilon^2 \left| \sum_{k=-K/2}^{K/2-1} X_m^*(k) H^*(k) e^{-j\frac{2\pi k}{N}n(1+\eta)} e^{-j\frac{2\pi k}{N}\eta N_m} \right|^2, \quad (21\text{a})$$

$$\Theta_{m,n} = -2\varepsilon\left(\frac{2\pi}{N}(N_m+n)\right)^2$$
$$\times \sum_{k=-K/2}^{K/2-1}\sum_{k'=-K/2}^{K/2-1} X_m(k')X_m^*(k)k'H(k')H^*(k)e^{-j\frac{2\pi(k-k')}{N}(n+\eta n+\eta N_m)}, \quad (21b)$$

$$\Pi_{m,n} = -\left(\frac{2\pi}{N}(N_m+n)\right)^2\left|\sum_{k=-K/2}^{K/2-1} kX_m(k)H(k)e^{j\frac{2\pi k}{N}n}e^{j\frac{2\pi k}{N}\eta(N_m+n)}\right|^2. \quad (21c)$$

The results derived in [4, Eqs. (28) and (29)] should be changed to the equations (20b) and (21b), respectively. The CRBs of the estimated parameters can then be obtained by the inversion $\mathbf{F}^{-1}$, which is given by

$$CRB_\varepsilon = \frac{F_{1,1}}{F_{0,0}F_{1,1} - F_{1,0}F_{0,1}}, \quad (22)$$

and

$$CRB_\eta = \frac{F_{0,0}}{F_{0,0}F_{1,1} - F_{1,0}F_{0,1}}. \quad (23)$$

**5. Simulation Results**

We now investigate the performance of the proposed ML estimator by Monte Carlo simulations. In the simulations, QPSK-modulated training symbols are used. The OFDM system parameters are based on IEEE 802.11a uncoded systems [6], where the DFT size $N$, the number of modulated subcarriers $K$, and the CP size $N_g$ are 64, 52, and 16, respectively. We consider frequency-selective fading channels with an exponential power delay profile given by $E\left[|h_l|^2\right] = e^{-l/5}\bigg/\sum_{l=0}^{L-1} e^{-l/5}$, where $l = 0, \cdots, L-1$ and $L = 5$. Each multipath is modeled as a zero-mean complex Gaussian random variable so that it varies according to the Rayleigh distribution [4]. In order to calculate joint ML estimates, we perform exhaustive searches of (11) or (12) by sufficiently quantizing the possible CFO $\varepsilon = i/100$ for $i \in \{-50, \cdots, -1, 0, 1, \cdots, 50\}$ and the possible SFO $\eta = i/100000$ for $i \in \{-50, \cdots, -1, 0, 1, \cdots, 50\}$.

Fig. 2 compares the mean square error (MSE) performances between Nguyen-Le's ML estimator [4] and the proposed ML scheme, where the MSEs of CFO and SFO estimates are defined as $MSE_\varepsilon = E\left[|\hat{\varepsilon} - \varepsilon|^2\right]$ and $MSE_\eta = E\left[|\hat{\eta} - \eta|^2\right]$, respectively. In the simulations, the normalized CFO $\varepsilon$ and SFO $\eta$ are 0.212 and 0.000112, respectively. It is seen that the proposed ML estimator outperforms Nguyen-Le's ML estimator at all SNR ranges for both CFO and SFO. The CRBs using (22) or (23) are also presented for comparison. The MSEs of the proposed ML scheme cannot achieve the CRBs because the proposed scheme uses frequency-domain received signals while the CRBs exploit time-domain received signals.

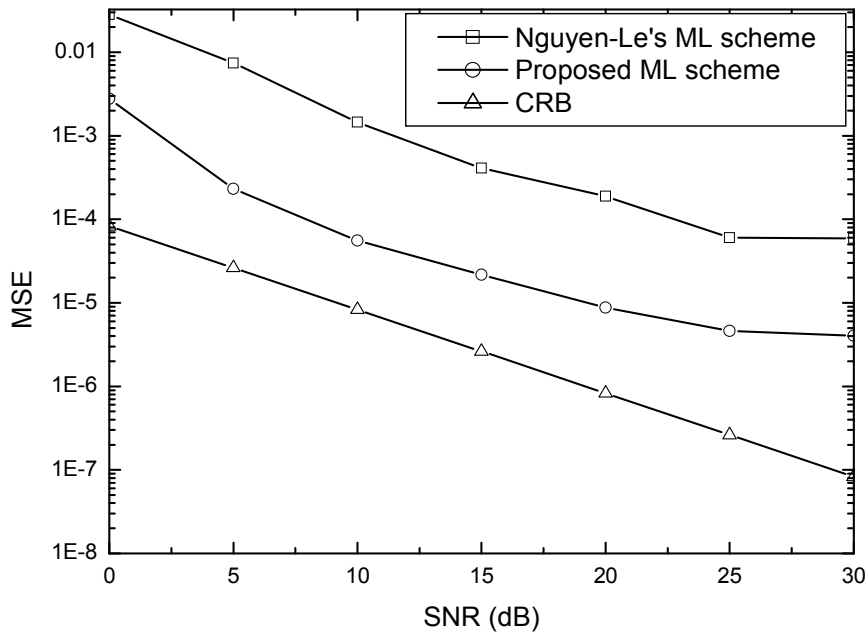

(a)

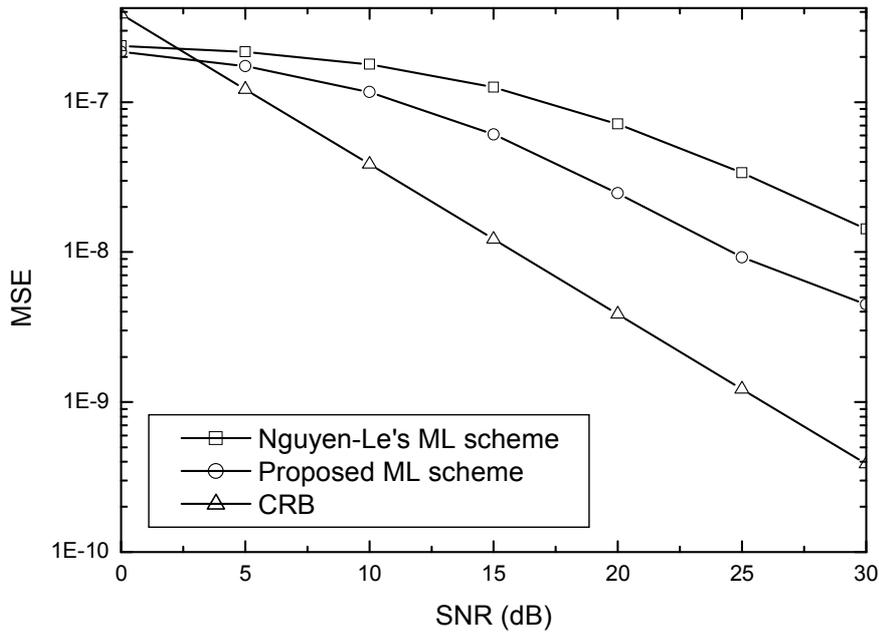

(b)

Fig. 2. MSEs and CRBs of CFO and SFO estimates versus SNR in a Rayleigh fading channel (CFO $\varepsilon = 0.212$ and SFO $\eta = 0.000112$): (a) CFO estimation; (b) SFO estimation

## 6. Conclusion

We have extended Moose's ML estimation to joint ML estimation of both CFO and SFO in OFDM systems. It is shown that the proposed ML estimation algorithm is superior in performance to Nguyen-Le's ML estimation algorithm. Therefore, the proposed ML estimator can be used for initial CFO and SFO estimation in the RLS-based iterative estimation and tracking algorithm proposed by Nguyen-Le et al.